\documentstyle[aps,twocolumn,epsf]{revtex}

\begin{document}

\title{Experimental Proposal for Achieving Superadditive Communication Capacities\\
       with a Binary Quantum Alphabet}
\author{J. R. Buck, S. J. van Enk, and Christopher A. Fuchs}
\address{Norman Bridge Laboratory of Physics 12-33, California
         Institute of Technology, Pasadena, CA  91125}
\date{5 March 1999}
\maketitle

\begin{abstract}
 We demonstrate superadditivity in the communication capacity of
a binary alphabet consisting of two nonorthogonal quantum states.
For this scheme, collective decoding is performed two
transmissions at a time.  This improves upon the previous schemes
of Sasaki {\em et al}.\ [Phys.\ Rev.\ A {\bf 58}, 146 (1998)]
where superadditivity was not achieved until a decoding of three
or more transmissions at a time.  This places superadditivity
within the regime of a near-term laboratory demonstration.  We
propose an experimental test based upon an alphabet of low
photon-number coherent states where the signal decoding is done
with atomic state measurements on a single atom in a high-finesse
optical cavity.
\end{abstract}

\section{Introduction}

This paper is about achieving the maximal information transfer
rate possible when information is encoded into quantum systems via
the preparation of one or another of two nonorthogonal states.
This might at first seem like a questionable thing to consider:
for transmissions through a noiseless medium, the maximal transfer
rate (or capacity) of 1 bit/transmission is clearly achieved {\it
only\/} with orthogonal alphabets.  This is because nonorthogonal
preparations cannot be identified with complete reliability.
However, there are instances in which it is neither practical nor
desirable to use such an alphabet.  The most obvious example is
when a simple laser transmitter is located a great distance from
the receiver.  The receiver's field will take on the character of
a very attenuated optical coherent state.  Because the states
become less orthogonal as the power is attenuated, one is
confronted with precisely the issue considered here. In this case,
one is typically stuck with trying to extract information from
quantum states that are not only nonorthogonal, but almost
completely overlapping.

The practical method in many situations for compensating for very
weak signals is to invest in elaborate receiving stations.  For
instance, in microwave communication very large-dish antennas are
the obvious route.  Recently, however, a new quantum mechanical
effect has been discovered for the decoding of nonorthogonal
signals on separate quantum systems.  Traditional signal
processing methods have only considered fixed decoding
measurements performed on the separate transmissions (see for
example Ref.~\cite{caves}): i.e., taking into account the
intrinsic noise generated by the quantum measurement
\cite{Holevo73}, one is left with a basic problem of classical
information theory---coding for a discrete memoryless channel
\cite{Cover91}. Quantum mechanics, however, allows for more
possibilities than this \cite{Holevo77}. If one is capable of
doing collective measurements on blocks of transmitted signals, it
is possible to achieve a greater capacity than one might have
otherwise thought \cite{Holevo79}---this is referred to as the
superadditivity of quantum channel capacities. This is an effect
that does not exist classically \cite[Lemma 8.9.2]{Cover91}.  The
physics behind the effect relies on a kind of nonlocality dual to
the famous one exhibited by entangled quantum systems through Bell
inequality violations \cite{Peres91,Bennett99,Fuchs99}.

More precisely, a communication rate $R$ is said to be achievable
if in $k$ transmissions there is a way of writing $2^{Rk}$
messages with the nonorthogonal alphabet so that the probability
of a decoding error goes to zero as $k\rightarrow\infty$.  The
number $R$ signifies the number of bits per transmission that can
be conveyed reliably from the transmitter to the receiver in the
asymptotic limit. Clearly the rates that can be achieved will
depend on the class of codings used for the messages and the class
of quantum measurements allowed at the receiver.  The capacity
$C_{n}$ is defined to be the supremum of all achievable rates,
where $n$ is the number of transmissions to be saved up before
performing a measurement.  The meaning of superadditivity is
simply that $C_n>C_1$, where the inequality is strict.

Generally it is a difficult problem to calculate $C_n$ even with a
quantum version of Shannon's noisy channel coding theorem
available \cite{Holevo79}.  And it is a much more difficult task
still to find codes that approach $C_n$.  This is because the
coding theorems generally give no information on how to construct
codes that approach a given capacity.  It turns out however that
the number $C_1$ is rather easily calculable and, because of a
recent very powerful theorem on quantum channel capacities
\cite{Hausladen96,Holevo98,Schumacher97}, so is the asymptotic
case $C_\infty$ \cite{Fuchs99}.  The most striking thing about
these two quantities is that even though both $C_1\rightarrow 0$
and $C_\infty\rightarrow 0$ as the overlap between the states goes
to unity, the ratio $C_{\infty}/C_{1}$ nevertheless diverges. This
means that grossly collective measurements can, in principle at
least, produce an arbitrarily large improvement in the channel
capacity of very weak signals---a very desirable state of affairs
and one of some serious practical import.

The problem from the practical side of the matter is that before
one will be able to decode very large blocks, one must first be
able to tackle the case of small blocks, preferably of just size
two or three.  There has already been substantial progress in this
direction by Sasaki {\em et al}., in a series of papers
\cite{Sasaki97,Sasaki98a,Sasaki98b}.  They explicitly demonstrate
a code that uses collective decoding three transmissions at a time
to achieve a communication rate $R_3$ greater than $C_1$.
Nevertheless, it would be nice to demonstrate superadditivity with
an even simpler scheme, namely two-shot collective measurements.
Also the ratio $R_3/C_1\rightarrow 1$ as the angle $\gamma$
between the two states goes to zero for their given coding scheme.
Thus just where one would be looking for the most help from
superadditivity (in the very weak signal regime), one loses it for
this code.

We improve on the work of Sasaki {\em et al}., by showing that in
fact $C_{2} > C_{1}$ for angles $\gamma\lesssim 19^{\circ}$, and
moreover that this superadditivity is sustained and only
strengthened as $\gamma\rightarrow 0$.   On the down side, the
improvement in capacity is not great---only 2.82 percent---but is
definitely there and not so small as to be forever invisible. In
this vein, we propose an experimental demonstration that relies on
near-term laboratory capabilities for implementation. For our two
nonorthogonal quantum states, we use low photon-number coherent
states $|\alpha\rangle$ and $|-\alpha\rangle$ with the separate
signals carried on different circular polarizations. The two-shot
signal decoding is performed with atomic state measurements on a
single Cesium atom in a high-finesse optical cavity via the
technique of quantum jumps in fluorescence similar to those
demonstrated on ions in
Refs.~\cite{Nagourney86,Sauter86,Bergquist86}.

The remainder of the paper is organized as follows.  In the next
Section, we demonstrate explicitly that $C_2>C_1$ for an alphabet
of two nonorthogonal states and compare that to the rate $R_3$
found in Refs.~\cite{Sasaki97,Sasaki98a,Sasaki98b}.  In Section
III, we specialize this calculation to the coherent states
mentioned above and approximate the decoding scheme of Section II
with a related scheme that is first order in the small parameter
$\alpha$.  Finally, in Section IV, we delineate the details of our
experimental proposal.

\section{Deriving Superadditivity for Two-Shot Collective
Measurements}

Following the discussion in the Introduction, we will take as an
alphabet for all communication schemes a fixed set of two
nonorthogonal quantum states $|\psi_0\rangle$ and $|\psi_1\rangle$
characterized by the single parameter $\gamma$:
\begin{equation}
\langle\psi_{0}|\psi_{1}\rangle=\cos\gamma\;.
\end{equation}
We would like to know what communication rates $R_n$ can be
achieved with this alphabet when decoding measurements are
performed $n$ transmissions at a time.

This in general is a very difficult problem, especially if one is
also confronted with the issue of explicitly demonstrating codes
for achieving those rates.  However, if one can be contented in
knowing the number $C_n$ itself and the quantum measurements
required to achieve that (i.e., without knowing the coding scheme
explicitly), then a great simplification arises because of a
quantum extension to the Shannon noisy coding theorem
\cite{Cover91} due to Holevo \cite{Holevo79}.

We shall state the result of this theorem presently.  Let the
variable $x$ denote the binary strings of length $n$ that index
the set of all messages
$|\Psi_x\rangle=|\psi_{x_1}\rangle|\psi_{x_2}\rangle\cdots
|\psi_{x_n}\rangle$, let the function $p(x)$ denote a probability
distribution over those messages, and let
\begin{equation}
\rho=\sum_x p(x)|\Psi_x\rangle\langle\Psi_x|
\end{equation}
denote the resultant density operator of that distribution.  We
shall use the notation $E$ to denote a generalized quantum
measurement or positive operator-valued measure (POVM)
\cite{Peres93} on the message Hilbert space ${\cal H}_n$, i.e.,
$E=(E_k)$ is an infinite sequence of operators on ${\cal H}_n$
with only a finite number of $E_k\ne0$ such that
$\langle\psi|E_k|\psi\rangle\ge0$ for all $k$ and $|\psi\rangle$,
and the $E_k$'s form a decomposition of the identity operator on
${\cal H}_n$.  In order to find $C_n$, it is enough to perform the
following maximization:
\begin{equation}
C_n = \frac{1}{n} \max_{p(x)}\, \max_{\rm E}\left[
H_{\rm\scriptscriptstyle E}(\rho) - \sum_x p(x)
H_{\rm\scriptscriptstyle E}(|\Psi_x\rangle)\right]\;,
\label{Candy-O}
\end{equation}
where
\begin{equation}
H_{\rm\scriptscriptstyle E}(\rho) = -\sum_k ({\rm tr}\rho E_k)\log
({\rm tr}\rho E_k)
\end{equation}
and
\begin{equation}
H_{\rm\scriptscriptstyle E}(|\Psi_x\rangle) = -\sum_k
\langle\Psi_x|E_k|\Psi_x\rangle\log\langle\Psi_x|E_k|\Psi_x\rangle
\end{equation}
are the Shannon informations for the various probability
distributions generated by the measurement $E$.  (In these
expressions we have used the base-two logarithm so that
information is measured in bits.)  For any rate
$R_n=C_n-\epsilon$, $\epsilon>0$, there {\it exists\/} a code that
will achieve that rate.  Moreover, if $E$ is fixed and only the
maximization over $p(x)$ is performed in Eq.~(\ref{Candy-O}), then
the resulting expression will define the capacity that can be
reached with the given measurement.

There are two limiting cases where the calculation of $C_n$
becomes tractable, $n=1$ and $n=\infty$.  In the first case, one
can use Refs.~\cite{Levitin95,Fuchs94,Fuchs99} to find rather
easily that
\begin{eqnarray}
C_{1}(\gamma) = &&
\frac{1}{2}\left[1+\sqrt{1-\cos^{2}\gamma}\,\right]\log
\left[1+\sqrt{1-\cos^{2}\gamma}\,\right] \nonumber\\ &&
+\frac{1}{2}\left[1-\sqrt{1-\cos^{2}\gamma}\,\right]\log
\left[1-\sqrt{1-\cos^{2}\gamma}\,\right]\;. \label{LittleBoy}
\end{eqnarray}
For the limit where arbitrarily many collective measurements are
made, one can use the powerful theorem of Ref.~\cite{Hausladen96}
to find that the channel capacity per bit is given by
\cite{Fuchs99}
\begin{eqnarray}
C_{\infty}(\gamma)= &-& \frac{1}{2}(1-\cos\gamma)\log\frac{1}{2}
(1-\cos\gamma) \nonumber\\ &-& \frac{1}{2}(1+\cos\gamma)\log
\frac{1}{2}(1+\cos\gamma)\;. \label{FatMan}
\end{eqnarray}
For all cases in between, there is nothing better to be done than
an explicit search over all probabilities $p(x)$ and all
measurements $E$.

As stated in the Introduction, one can see from
Eqs.~(\ref{LittleBoy}) and (\ref{FatMan}), that
\begin{equation}
\lim_{\gamma \rightarrow 0}
\frac{C_\infty(\gamma)}{C_1(\gamma)}\;\;\longrightarrow\;\;\infty\;.
\end{equation}
So the incentive to use collective measurements in the decoding of
these signals is great.

Therefore, let us specialize to the case of collective
measurements on two transmissions at a time.  In this case, with
respect to the decoding observables we have an effective alphabet
consisting of the tensor-product states
\begin{eqnarray}
|a\rangle &=& |\psi_{0}\rangle |\psi_{1}\rangle\\ |b\rangle &=&
|\psi_{1}\rangle |\psi_{0}\rangle\\ |c\rangle &=& |\psi_{0}\rangle
|\psi_{0}\rangle\\ |d\rangle &=& |\psi_{1}\rangle
|\psi_{1}\rangle\;,
\end{eqnarray}
with the consequent inner products
\begin{equation}
\langle a|c\rangle=\langle b|c\rangle = \langle
 a|d\rangle = \langle b|d\rangle = \cos\gamma\;,
\end{equation}
and
\begin{equation}
\langle a|b\rangle=\cos^2\gamma\;.
\end{equation}

It turns out that these states can already exhibit superadditivity
even when the collective observables are taken to be simple von
Neumann measurements: i.e., by taking $E_k=|e_k\rangle\langle
e_k|$ where the $|e_k\rangle$ are four orthonormal vectors. Taking
$p_{i}$ to be a probability distribution on the effective alphabet
states, we must attempt to maximize the rate

\begin{figure}[htbp]
\label{fourd}
  \begin{center}
   \leavevmode
     \epsfxsize=7cm  \epsfbox{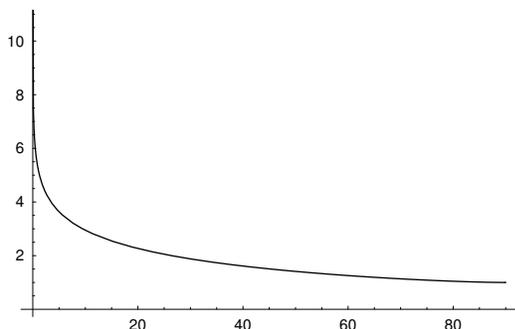}
\caption{The ratio $C_{\infty}/C_{1}$ as a function of the angle
$\gamma$ in degrees, where $\gamma$ is the angle between the two
nonorthogonal states comprising the transmission alphabet.}
  \end{center}
\end{figure}

\begin{eqnarray}
R = && H_{\rm\scriptscriptstyle E}(\rho) -
p_{a}H_{\rm\scriptscriptstyle E}(|a\rangle) -
p_{b}H_{\rm\scriptscriptstyle E}(|b\rangle) -
p_{c}H_{\rm\scriptscriptstyle E}(|c\rangle) \nonumber\\ && -
p_{d}H_{\rm\scriptscriptstyle E}(|d\rangle)
\end{eqnarray}
with
\begin{equation}
\rho=p_{a}|a\rangle\langle a|+p_{b}|b\rangle\langle b|+
p_{c}|c\rangle\langle c|+p_{d}|d\rangle\langle d|\;,
\end{equation}
\begin{equation}
H_{\rm\scriptscriptstyle E}(\rho)=-\sum_{k} \langle
e_{k}|\rho|e_{k}\rangle\log \langle e_{k}|\rho|e_{k}\rangle\;,
\end{equation}
and
\begin{equation}
H_{\rm\scriptscriptstyle E}(|a\rangle)= -\sum_{k}|\langle
e_{k}|a\rangle |^{2}\log|\langle e_{k}|a\rangle|^{2}
\end{equation}
and likewise for $|b\rangle$, $|c\rangle$, and $|d\rangle$. The
rate $R_{2}$ we will be interested in is then
\begin{equation}
R_{2}=\frac{1}{2}\max_{p_{i}}\,\max_{|e_k\rangle} R\;.
\label{Flubber}
\end{equation}

We have thoroughly studied $R_2$ numerically with a steepest
descent and simulated annealing technique.  As one might guess,
the optimal solution to Eq.~(\ref{Flubber})---for sufficiently
small angles ($\gamma\lesssim 19^{\circ}$)---appears to obey the
following symmetries
\begin{eqnarray}
p_{d}&\rightarrow& 0,\nonumber\\
    p_{a}&=&p_{b}\equiv p,\nonumber\\
    p_{c} &\equiv& 1 - 2p.
\end{eqnarray}
Therefore we make the following {\em Ansatz\/} (see Fig.~2)
\begin{eqnarray} \langle c|e_{3}\rangle&=&\cos\eta\nonumber\\
    \langle a|e_{1}\rangle &=& \langle b|e_{2}\rangle\nonumber\\
    \langle a|e_{3}\rangle &=& \langle b|e_{3}\rangle\nonumber\\
    \langle c|e_{1}\rangle &=& \langle c|e_{2}\rangle
\end{eqnarray}

\begin{figure}[htbp]
\label{fourc}
  \begin{center}
   \leavevmode
     \epsfxsize=7cm  \epsfbox{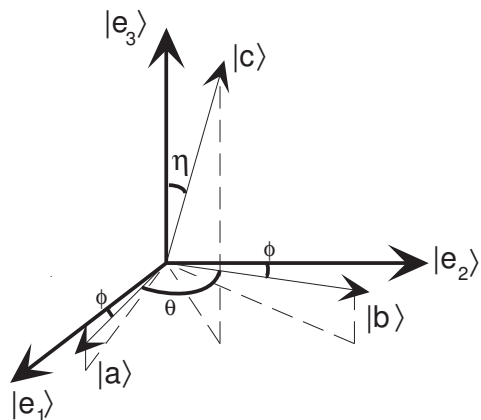}
\caption{The effective alphabet for our implementation represented
in an orthogonal measurement basis. The projections are in the
$|e_{1}\rangle$,$|e_{2}\rangle$ plane.}
  \end{center}
\end{figure}

\begin{figure}[htbp]
\label{fourb}
  \begin{center}
   \leavevmode
     \epsfxsize=7cm  \epsfbox{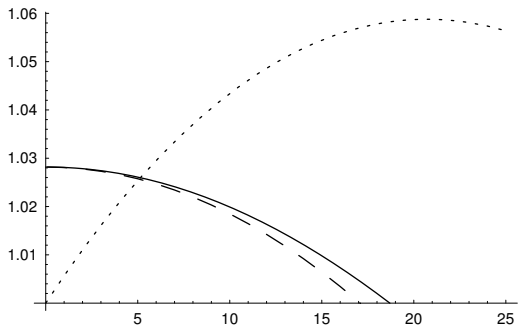}
\caption{The solid line represents the ratio $R_{2}/C_{1}$ as a
function of $\gamma$. The dashed line represents the ratio
obtained using the experimentally feasible but nonoptimal basis
discussed in Section III. The dotted line represents $R_{3}/C_{1}$
obtained by Sasaki {\em et al}. in
Refs.~{\protect\cite{Sasaki97,Sasaki98a,Sasaki98b}}.}
  \end{center}
\end{figure}

Taking these symmetries as a more analytic starting point, we can
expand the measurement basis as a function of $\eta$, $\gamma$,
and the alphabet states (see Fig.~2):
\begin{eqnarray}    |e_{1}\rangle =&&
\frac{\cos\eta + 1}{2 \sin\gamma}|a\rangle + \frac{\cos\eta - 1}{2
\sin\gamma}|b\rangle\nonumber\\
            &&+ \frac{\sqrt{2}\sin\eta\sin\gamma
            - 2\cos\eta\cos\gamma}{2\sin\gamma}|c\rangle\nonumber\\
    |e_{2}\rangle =&& \frac{\cos\eta - 1}{2 \sin\gamma}|a\rangle + \frac{\cos\eta + 1}{2
    \sin\gamma}|b\rangle\nonumber\\
            &&+ \frac{\sqrt{2}\sin\eta\sin\gamma
            - 2\cos\eta\cos\gamma}{2\sin\gamma}|c\rangle\nonumber\\
    |e_{3}\rangle =&& - \frac{\sqrt{2}\sin\eta}{2\sin\gamma}|a\rangle
        - \frac{\sqrt{2}\sin\eta}{2\sin\gamma}|b\rangle\nonumber\\
        &&+ \frac{\sqrt{2}\sin\eta\cos\gamma+\cos\eta\sin\gamma}{\sin\gamma}|c\rangle
\end{eqnarray}
Thus the rate can now be expressed as
\begin{equation} R_{2}(\gamma) = \max_{\eta,p}R(\eta,p,\gamma)
\end{equation}

\begin{figure}[htbp]
\label{foura}
  \begin{center}
   \leavevmode
     \epsfxsize=7cm  \epsfbox{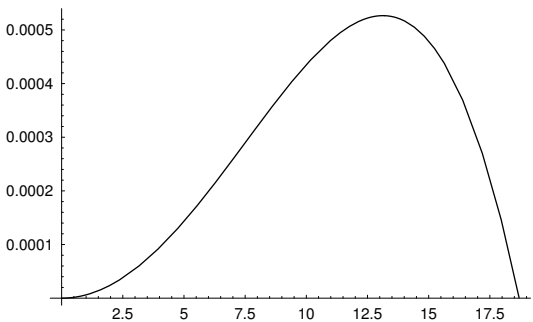}
\caption{The difference in rates $R_{2}-C_{1}$ as a function of
the angle $\gamma$.}
  \end{center}
\end{figure}

Even with these strong assumptions and simplifications,
$R_2(\gamma)$ does not yield a simple analytic expression.  We
must instead content ourselves with a numerical study as depicted
in Figure 3.  Note in particular that as $\gamma\rightarrow0$ the
superadditivity does not dwindle away:
\begin{equation}
\lim_{\gamma \rightarrow 0} \frac{R_{2}(\gamma)}{C_{1}(\gamma)}
\;\;\longrightarrow\;\; 1.02818\;.
\end{equation}
This contrasts with the rate $R_3$ exhibited by Sasaki {\em et
al}.\ \cite{Sasaki97,Sasaki98a,Sasaki98b} for which the ratio
$R_3/C_1$ goes to one within the very weak signal regime.

Note that we use the notation $R_{2}$ rather than $C_{2}$ because
our favored quantity can only be asserted as a {\it lower bound\/}
to the two-shot capacity.  The symmetry assumptions on the
probabilities along with the specialization to symmetric von
Neumann measurements could turn out to be overly restrictive.
However, further numerical investigations seem to indicate that
any further improvement is likely to be very small
\cite{SmolinPrivates}. Also we should emphasize that demonstrating
that $R_{2}>C_1$ does not give an automatic means for finding a
code that comes within $\epsilon$ of this rate:  the channel
capacity theorem Eq.~(\ref{Candy-O}) is only an existence proof of
such a code. However, the noise model that our alphabet and
measurement leads to---i.e., a simple stochastic transition
diagram on three letters---has been extensively studied in
classical information-theory literature, and good codes for this
problem are likely to exist.

Finally, let us mention one more quantification of the
superadditivity due to our nonorthogonal alphabet; this is the
simple difference between the two-shot rate and the one-shot rate.
We plot $R_{2}-C_{1}$ in Figure 4.  It has been suggested in
Ref.~\cite{Fuchs99} that the differences $C_n-C_1$ can help define
various notions of when two quantum states are most ``quantum''
with respect to each other (and hence least ``classical'').  When
one goes to the limit $C_\infty-C_1$ one finds a well-behaved
notion: two states are most quantum with respect to each other
when they are $45^\circ$ apart.  Figure 4 seems to indicate that
$R_{2}-C_{1}$ plays no such simple role:  at the very least, it
means that this difference does a poor job of ferreting out the
quantumness of two states in the geometrical sense already
supplied by Hilbert space.

\section{Basis for Experiment}

Let us now focus on the case we are most interested in for our
experimental proposal:  two very low photon-number coherent states
$|\alpha\rangle$ and $|-\alpha\rangle$ of a particular field mode.
We choose $\alpha$ real so that the mean photon number in that
mode is $\alpha^2$. For the angles for which we demonstrated
superadditivity, i.e., $\gamma\lesssim 19^\circ$, this translates
to a mean photon number less than 0.03 in each transmission.  In
this case, we are well warranted in making the approximation
\begin{eqnarray}\label{sukkel}
|\psi_{0}\rangle &=&
|\alpha\rangle\cong\frac{1}{\sqrt{1+\alpha^{2}}}|0\rangle +
\frac{\alpha}{\sqrt{1+\alpha^{2}}}|1\rangle \nonumber\\
|\psi_{1}\rangle &=& |-
\alpha\rangle\cong\frac{1}{\sqrt{1+\alpha^{2}}}|0\rangle -
\frac{\alpha}{\sqrt{1+\alpha^{2}}}|1\rangle\;,
\end{eqnarray}
where $|0\rangle$ and $|1\rangle$ denote the zero- and
single-photon states of the mode, respectively.  Moreover, we have
\begin{equation}
\alpha \cong \sqrt{\frac{1-\cos\gamma}{1+\cos\gamma}}\;.
\end{equation}
In order to keep track of the separate transmissions, we encode
each transmission in a different mode. For our purposes it is
convenient to choose two orthogonal circular
polarizations.\footnote{Perhaps superfluously, we note that one
might have imagined meeting the power constraint with an alphabet
of two coherent states of identical amplitude $\alpha$ but of
different polarization modes. Such an alphabet takes the form
 $|\psi_0\rangle=|\alpha\rangle_{+}|0\rangle_-$
and $|\psi_1\rangle=|0\rangle_+|\alpha\rangle_{-}$, where the
tensor product structure now reflects the fact that one is using
two field modes for each single transmission. However, this
alphabet of states is even less orthogonal than the one defined in
Eq.~(\ref{sukkel}).} Expanding the measurement basis in terms of
the photon number states, we thus have
\begin{eqnarray}
|e_{1}\rangle =&&
\frac{\sqrt{2}\sin\eta+2\alpha\cos\eta}{2(1+\alpha^{2})}
|0\rangle_{+}|0\rangle_{-} \nonumber\\ &&+
\frac{\alpha\sqrt{2}\sin\eta-\cos\eta+\alpha^{2}
\cos\eta-1-\alpha^{2}}{2(1+\alpha^{2})}|0\rangle_{+}|1\rangle_{-}
\nonumber\\ &&+ \frac{\alpha\sqrt{2}\sin\eta-\cos\eta+\alpha^{2}
\cos\eta+1+\alpha^{2}}{2(1+\alpha^{2})}|1\rangle_{+}|0\rangle_{-}
\nonumber\\ &&+
\frac{\alpha^{2}\sqrt{2}\sin\eta-2\alpha\cos\eta}{2(1+\alpha^{2})}
|1\rangle_{+}|1\rangle_{-}
\\
\nonumber\\ |e_{2}\rangle =&&
\frac{\sqrt{2}\sin\eta+2\alpha\cos\eta}{2(1+\alpha^{2})}
|0\rangle_{+}|0\rangle_{-} \nonumber\\ &&+
\frac{\alpha\sqrt{2}\sin\eta-\cos\eta+\alpha^{2}
\cos\eta+1+\alpha^{2}}{2(1+\alpha^{2})}|0\rangle_{+}|1\rangle_{-}
\nonumber\\ &&+ \frac{\alpha\sqrt{2}\sin\eta-\cos\eta+\alpha^{2}
\cos\eta-1-\alpha^{2}}{2(1+\alpha^{2})}|1\rangle_{+}|0\rangle_{-}
\nonumber\\ &&+
\frac{\alpha^{2}\sqrt{2}\sin\eta-2\alpha\cos\eta}{2(1+\alpha^{2})}
|1\rangle_{+}|1\rangle_{-}
\\
\nonumber\\ |e_{3}\rangle =&&
\frac{\cos\eta-\alpha\sqrt{2}\sin\eta}{(1+\alpha^{2})}|0\rangle_{+}
|0\rangle_{-} \nonumber\\ &&+
\frac{\sqrt{2}\sin\eta(1-\alpha^{2})+
2\alpha\cos\eta}{2(1+\alpha^{2})}|0\rangle_{+}|1\rangle_{-}
\nonumber\\ &&+ \frac{\sqrt{2}\sin\eta(1-\alpha^{2})+
2\alpha\cos\eta}{2(1+\alpha^{2})}|1\rangle_{+}|0\rangle_{-}
\nonumber\\ &&+
\frac{\alpha\sqrt{2}\sin\eta+\alpha^{2}\cos\eta}{(1+\alpha^{2})}
|1\rangle_{+}|1\rangle_{-}
\end{eqnarray}
The $+$ and $-$ subscripts in these equations refer to righthand
and lefthand circularly polarized light, respectively.

The measurement basis above is, of course, orthonormal.  However,
after optimizing over $\eta$ as in the previous section, one finds
that the coefficient of each $|1\rangle_+|1\rangle_-$ component
turns out to be of order $\alpha$ while the other terms are of
order one. Because one is free to choose any measurement basis, we
choose to ignore the small $|1\rangle_+|1\rangle_-$ term for each
$|e_i\rangle$. This new basis $|\tilde{e}_i\rangle$ is close to
the optimal basis $|e_i\rangle$ but allows the great
simplification of not having to worry about how to distinguish
$|1\rangle_+|1\rangle_-$ from $|0\rangle_+|1\rangle_-$ and
$|1\rangle_+|0\rangle_-$.  We may then focus on experiments based
on the absorption of at most a single photon.

The final step for defining our measurement scheme is to
re-orthogonalize the vectors $|\tilde{e}_i\rangle$. A simple
convenient technique for this is the one introduced in
Ref.~\cite{Hughston93}.  Let
 \begin{equation}
M = \sum_{i=1}^{3}|\tilde{e}_i\rangle\langle\tilde{e}_i|\;.
\end{equation}
Then clearly the vectors
\begin{equation}
|e^\prime_{i}\rangle = M^{-\frac{1}{2}}|\tilde{e}_i\rangle
\label{Mambo}
\end{equation}
form an orthonormal set.  It is this basis that we will use in the
experimental proposal, the main point of interest about it being
that it contains no two-photon contributions.  Of course, the new
basis cannot be optimal for achieving the rate $R_2$ already
calculated, but for small $\alpha$ it becomes arbitrarily good. In
fact, it is already sufficient for demonstrating superadditivity
for $\gamma\lesssim 17^\circ$ (see Fig.~3).

\section{Experimental Proposal}

We now turn to the task of realizing the measurement explored in
the last section.  To carry this out, we need the ability to
perform an entangled measurement on two wave packets at a time. We
can achieve this collective decoding by mapping the orthonormal
measurement basis in Eq.~(\ref{Mambo}) onto a set of orthonormal
superpositions of three sublevels of a single atom (see Figure 5).

The basic idea is to first transfer the information from the
propagating light fields to photons inside an optical cavity and
subsequently map the information from the cavity field to a single
atom inside that cavity. This can be accomplished as follows:
First, the atom is prepared in a ground state with $|m=0\rangle$
by optical pumping. The presence of a single $\sigma^+$ polarized
cavity photon is then more than sufficient to induce a Raman
transition to the $|m=1\rangle$ state with the help of a
$\pi$-polarized laser field (in fact, the advances in cavity QED
have increased the

\begin{figure}[htbp]
\label{four}
  \begin{center}
   \leavevmode
     \epsfxsize=7cm  \epsfbox{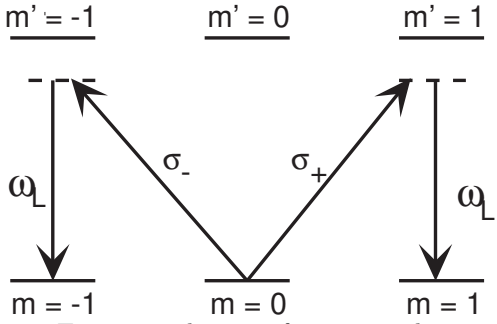}
\caption{Transition diagram for our implementation: a
$\pi$-polarized laser field with frequency $\omega_L$ is applied
to a single atom inside an optical cavity. The laser will induce a
Raman transition from the initial state $|m=0\rangle$ to
$|m=+1\rangle$ or $|m=-1\rangle$ in the presence of a single
$\sigma^+$ or $\sigma^-$ polarized cavity photon, with frequency
$\omega_C = \omega_L$. No transition is induced in the absence of
a cavity photon, as the $m=0\leftrightarrow m'=0$ transition is
forbidden. Note that $\pi$-polarized modes are not supported by
the cavity.}
\end{center}
\end{figure}

\noindent atom-cavity coupling to such a large degree that the
saturation photon number for optical transitions is very
small\cite{jeff}; in particular, for the
$(6S_{\frac{1}{2}},F=4,m=4)\rightarrow (6P_{\frac{3}{2}},F=5,m=5)$
transition in Cesium it is only $2.3\times 10^{-4}$\cite{hood}).
Similarly, the presence of a single $\sigma^-$ photon will induce
the transition to $|m=-1\rangle$, while if no cavity photon is
present, the atom will stay in $|m=0\rangle$. Thus, the
measurement scheme is based on the mapping

\begin{eqnarray}
|0\rangle_{+}|0\rangle_{-}|m=0\rangle&\;\;\longmapsto\;\;&|0\rangle_{+}
|0\rangle_{-}|m=0\rangle,\nonumber\\
|0\rangle_{+}|1\rangle_{-}|m=0\rangle&\;\;\longmapsto\;\;&|0\rangle_{+}
|0\rangle_{-}|m=-1\rangle,\nonumber\\
|1\rangle_{+}|0\rangle_{-}|m=0\rangle&\;\;\longmapsto\;\;&|0\rangle_{+}
|0\rangle_{-}|m=+1\rangle.
\end{eqnarray}
This mapping must be executed within the cavity lifetime (a
typical lifetime for high-finesse optical cavities is
$\kappa^{-1}\sim 0.1 \mu$s \cite{mabuchi}). Once this mapping has
been performed, we no longer rely on cavity fields.

In order to avoid any disturbing effects from the laser field on
level $|m=0\rangle$ in the absence of a cavity photon, we require
the transition $|m=0\rangle\mapsto |m'=0\rangle$ to be forbidden,
which is easily accomplished by choosing $\delta F=0$ transitions.
For example, one might consider the following transition between
hyperfine multiplets in Cesium
\begin{equation}
6S_{\frac{1}{2}},F=3
\;\;\longleftrightarrow\;\;6P_{\frac{1}{2}},F=3.
\end{equation}
Moreover, the frequency $\omega_L$ of the laser field is chosen
such that we are on two-photon resonance with the $|m=0\rangle
\leftrightarrow |m=\pm 1\rangle$ Raman transitions, but far off
resonance with respect to the excited states. Therefore, the
latter will not be populated and no further transitions from
$|m=1\rangle$ or $|m=-1\rangle$ will occur.

Once the information has thus been transferred from the
polarizations to the atom in the cavity, the measurement basis is
an orthonormal {\it superposition\/} of the three relevant atomic
ground states $|m=-1\rangle$, $|m=0\rangle$, and $|m=1\rangle$.
Making a measurement of a superposition of these states is far
more difficult than measuring the states themselves. Therefore, we
first apply a unitary operation that transforms the basis of
Eq.~(\ref{Mambo}) into the physical measurement basis. This
operation can be performed by a series of at most 16 appropriately
timed Raman pulses \cite{law,harel}. In general, for an $N$ level
system, with $N$ even, the unitary evolution can be controlled
with a sequence of $N^2$ pulses consisting of two distinct
perturbations in an alternating sequence\cite{harel}. While this
scheme is not optimal for $N=3$, it does give an upper bound for
the required number of pulses.

Once this transformation of basis has been performed, the only
remaining task is to measure the projection onto each of the three
possible hyperfine levels of our physical measurement basis. To
perform this measurement, a magnetic field is turned on
adiabatically, causing a splitting of the energy of these
otherwise degenerate hyperfine levels. Next, we use the technique
of optical shelving to make a measurement of the levels
\cite{Nagourney86,Sauter86,Bergquist86}. With this technique, a
Raman pulse is applied to cause a transition from the
$|m=1\rangle$ state into a secondary state that can then be driven
on resonance to yield a large number of photons. If the atom
fluoresces at the driven frequency, the measurement outcome is
$m=1$, and the measurement is finished. Otherwise, if no
fluorescence is detected, the atom will not be affected by this
driving laser and the process is then repeated for the
$|m=0\rangle$ and $|m=-1\rangle$ states.

Finally, let us note that atoms can already be held in a cavity
for times exceeding $250 \mu$s \cite{mabuchi}, which is nearly
sufficient for the measurements and laser manipulations discussed
to be performed. In addition, further improvements on trapping
atoms in cavities will relax the conditions on timing. The ability
to hold single atoms in a cavity for a sufficient period of time
will open up a world of possibilities for the field of
communication\cite{cirac97,vanenk98,vanenk97}.

\section{Acknowledgements}

We thank H.~J. Kimble, R. Legere, H.~Mabuchi, M.~Sasaki, P.~W.
Shor, and J.~A. Smolin for helpful discussions.  We thank
A.~Landahl for a careful reading of the manuscript. This work was
supported by the QUIC Institute funded by DARPA via the ARO, by
the ONR, and by the NSF\@. CAF also acknowledges the support of a
Lee A. DuBridge Fellowship.


\end{document}